\documentclass[final,5p,times]{elsarticle}

\makeatletter
\def\ps@pprintTitle{%
    \let\@oddhead\@empty
    \let\@evenhead\@empty
    \def\@oddfoot{\footnotesize\itshape
         {} \hfill}%
    \let\@evenfoot\@oddfoot
    }
\makeatother

\usepackage{amssymb}
\usepackage{amsmath}
\usepackage{textcomp}
\usepackage{xcolor}
\usepackage{braket}
\usepackage[normalem]{ulem}

\journal{ }

\begin{document}

\begin{frontmatter}



\title{Hybrid Quantum Neural Network Advantage for Radar-Based Drone Detection and Classification in Low Signal-to-Noise Ratio} 



\author[first] {Aiswariya Sweety Malarvanan}
\affiliation[first]{organization={Department of Physics, University of Cologne},
            addressline={Albertus-Magnus-Platz}, 
            city={Cologne},
            postcode={50923}, 
            state={NRW},
            country={Germany}}

\begin{abstract}

In this paper, we investigate the performance of a Hybrid Quantum Neural Network (HQNN) and a comparable classical Convolution Neural Network (CNN) for detection and classification problem using a radar. Specifically, we take a fairly complex radar time-series model derived from electromagnetic theory, namely the Martin-Mulgrew model, that is used to simulate radar returns of objects with rotating blades, such as drones. We find that when that signal-to-noise ratio (SNR) is high, CNN outperforms the HQNN for detection and classification. However, in the low SNR regime (which is of greatest interest in practice) the  performance of HQNN is found to be superior to that of the CNN of a similar architecture.

\end{abstract}



\begin{keyword}

Drone Classification \sep Radar Detection  \sep Convolutional Neural Networks \sep Hybrid Quantum Neural Networks 



\end{keyword}

\end{frontmatter}




\section{Introduction}
\label{introduction}

Machine learning has been successfully applied to a wide variety of problems. Quantum Machine Learning (QML) is an emerging field that merges the principles of quantum computing with classical machine learning techniques. The aim is to take advantage of quantum phenomena to enhance the efficiency and capabilities of classical machine learning algorithms \cite{introQML} \cite{QML} \cite{QMLbook}. 

Hybrid quantum neural networks (HQNNs), which leverages classical deep learning architectures alongside quantum algorithms, such as Variational Quantum Circuits (VQCs), to process data, stands out as a significant research domain in QML \cite{Quantum_circuit_learning}  \cite{Hybrid-parallel-theory} \cite{a9}. 

The use of HQNNs has been explored for a diverse set of problems including drug response prediction \cite{a1}, detection of dementia \cite{a8}, multi-classification tasks \cite{a7} \cite{a2}, remote sensing imagery classification \cite{a3}, detection of cyberattacks \cite{a4}, gender recognition from facial images \cite{a5}, and for the classification of finance and MNIST data \cite{a6} \cite{Hybrid-parallel-model-used}. 

The widespread use of drones across various sectors has motivated the need for robust methods for their reliable detection and classification. A major motivation for this is that the statistical model of radar returns is complex, and in many cases analytical solutions are not available.  The use of machine learning and deep learning for this problem has been investigated in some prior work \cite{b3} \cite{b4} \cite{b5} \cite{divy_paper}. 

In this article, we use an HQNN with parallel quantum dense layers for the drone detection and classification problem. Previous work had investigated the detection and identification of drones using CNN \cite{drone_detection} \cite{divy_paper}. This work extends the analysis to HQNN and compares the performance of the HQNN with a CNN of a comparable architecture. Specifically, as in the prior work, we employ radar micro-Doppler signatures obtained from the Short-Time Fourier Transform (STFT) spectrograms of the time-series data of the radar signals reflected from the drones, simulated based on the Martin-Mulgrew (MM) model. We study the data in X-band radar simulation scenarios. We examine the binary classification of drones versus noise for radar detection. We also compare the performance of HQNN models with classical CNNs for the multi-class classification problem. We find an unambiguous gain from using HQNNs in the low-SNR regime, the regime of greatest practical interest. The HQNN is found to perform better for lower SNR compared to a CNN of a comparable architecture. The results strongly suggest that HQNNs should be explored for the application to radar time series analysis.

In Section \ref{sec:Bkgd}, we review the radar model used in the paper, including the time domain model, the frequency domain representation, and the proposed detection and classification methodology. 

In Section \ref{sec:SimulationSetup}, the data generation method is presented, and the neural architecture is discussed, including CNN and HQNN. The model training approach in presented in Section \ref{ssec:ModelTraining}. The results are presented in Section \ref{sec:Results}, followed by the conclusion in Section \ref{sec:Conclusion}.

\section{Background}\label{sec:Bkgd}

\subsection{Radar Signal Model: Time Domain Model}\label{ssec:RadarModel}


The Martin-Mulgrew (MM) model \cite{MM}, originally designed to analyze radar returns of aircraft propeller blades, can model radar return signals (in the form of complex time series of desired time duration) of any aerial object with rotating propellers, such as drones. This MM model is given as:

\begin{equation}
\begin{aligned}
\label{equation_psi_generalized}
\Psi(t) = & A_{r} e^{j \left( 2 \pi f_{c} t - \frac{4 \pi}{\lambda} \left(R + V_{rad} t\right)  \right)}\\
&\sum_{n=0}^{N-1} 
\left( \alpha + \beta \cos \left( \Omega_{n} \right) \right)
e^{-j \frac{L_{1} + L_{2}}{2} \gamma_{n}}\text{sinc} \left( \frac{L_{2} - L_{1}}{2}  \gamma_{n} \right)
\nonumber
\end{aligned}
\end{equation}

where

\begin{equation}
\begin{aligned}
\alpha &= \sin \left( |\theta| + \Phi_{p} \right) + \sin \left( |\theta| -\Phi_{p} \right)
\\
\beta &= \text{sign} \left( \theta \right) \left( \sin \left( |\theta| + \Phi_{p} \right) - \sin \left( |\theta| - \Phi_{p} \right) \right)
\\ 
\Omega_{n} &= 2 \pi \left(f_{rot} t + \frac{n}{N} \right)
\\
\gamma_{n} &= \frac{4 \pi}{\lambda} \cos \left( \theta \right) \sin \left( \Omega_{n} \right)
\end{aligned}
\end{equation}

\begin{itemize}\item[]
\item[] $A_{r}$ is a real, scaling factor,
\item[] $L_{1}$ is the distance of the blade roots from the centre of rotation,
\item[] $L_{2}$ is the distance of the blade tips from the center of rotation,
\item[] $N$ is the number of blades,
\item[] $R$ is the range of the target,
\item[] $V_{rad}$ is the radial velocity of the center of rotation with respect to the radar,
\item[] $\lambda$ is the wavelength of the transmitted radar signal,
\item[] $\theta$ is the angle between the plane of rotation and the line of sight from the radar to the center of rotation,
\item[] $f_{c}$ is the frequency of the transmitted radar signal,
\item[] $f_{rot}$ is the frequency of rotation of drone blades,
\item[] $t$ is the time duration of radar returns,
\item[] $\Phi_p$ is the pitch of the blades
\end{itemize}

The simulation parameters $N$, $L_{1}$, $L_{2}$, and $f_{rot}$ characterises drone blades/propellers while $\lambda$ and $f_{c}$ determine the type of radar. The position of the drone relative to the radar is defined by $\theta$, $\Phi_p$, $R$ and $V_{rad}$. The data generated using the MM model is discretized at a sampling frequency of $f_s$.

An important parameter of interest is the single pulse signal-to-noise ratio, which is defined as $10\log_{10}A_r^2/\sigma_n^2$, where $\sigma_n^2$ is the noise variance. Note that coherent processing of $N_{\text{pulses}}$ pulses will give an additional gain of $10\log_{10}N_\text{pulses}$. Also note that since the model focuses on the radar returns from the rotating blades, the SNR under consideration could be substantially lower than that of the body (e.g., rotating blades on a drone relative to the drone body). For simplicity, by SNR we mean single-pulse SNR, unless specified otherwise. 

In this work, radar returns of an X-band radar from five different types of commercially available drones (DJI Mavic Air 2, DJI Mavic Mini, DJI Matrice 300 RTK, DJI Phantom 4 and Parrot Disco) are simulated using the MM model. The simulated X-band radar has a wavelength $\lambda$ of 3 cm, a transmitting frequency $f_{c}$ of 10 GHz, and a pulse repetition frequency (PRF) of 10kHz. The approximate drone parameters of each drone for which the radar returns are simulated are shown in Table \ref{drone_characteristics}.

\begin{table}
\centering
\begin{tabular}{l c c c c} 
\hline
 Drone & N & $L_{1}$ [cm] & $L_{2}$ [cm] & $f_{r}$ [Hz] \\ 
\hline
 DJI Mavic Air 2     & 2 & 0.50 & 7.00  & 91.66 \\ 
 DJI Mavic Mini 2    & 2 & 0.50 & 3.50  & 160.00 \\ 
 DJI Matrice 300 RTK & 2 & 5.00 & 26.65 & 70.00 \\ 
 DJI Phantom 4       & 2 & 0.60 & 5.00  & 116.00 \\ 
 Parrot Disco        & 2 & 1.00 & 10.40 & 40.00 \\
 \hline
\end{tabular}
\caption{Approximate drone parameters of five drones used in this study}
\label{drone_characteristics}
\end{table}

\subsection{Frequency Domain Representation}\label{ssec:FrequencyDomain}

The radar model time series is time-varying and nonstationary. As a result, frequency is constant and well defined only in a short time window. Hence, the short-time Fourier transform (STFT) is utilized to obtain the frequency spectrum. 

\begin{figure}
	\centering 
	\includegraphics[width= .5 \textwidth, angle=0]{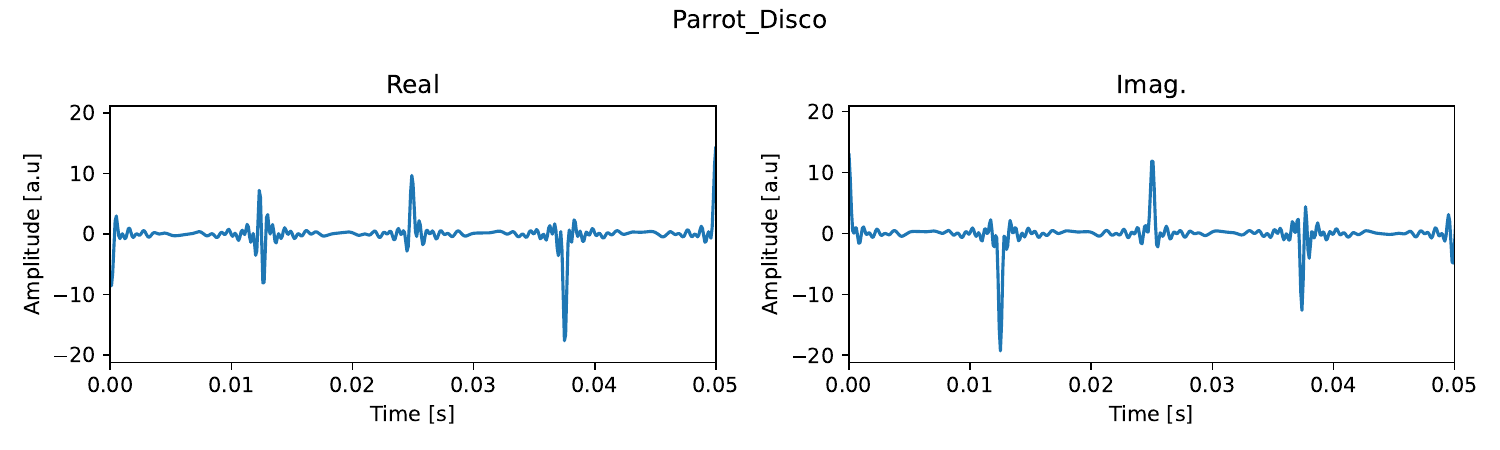}	
	\caption{Real and imaginary valued portions of the time series for drone Parrot Disco, produced using the MM model. The signal contains no noise.} 
	\label{time_series_drone}%
\end{figure}

\begin{figure}
	\centering 
	\includegraphics[width= .5 \textwidth, angle=0]{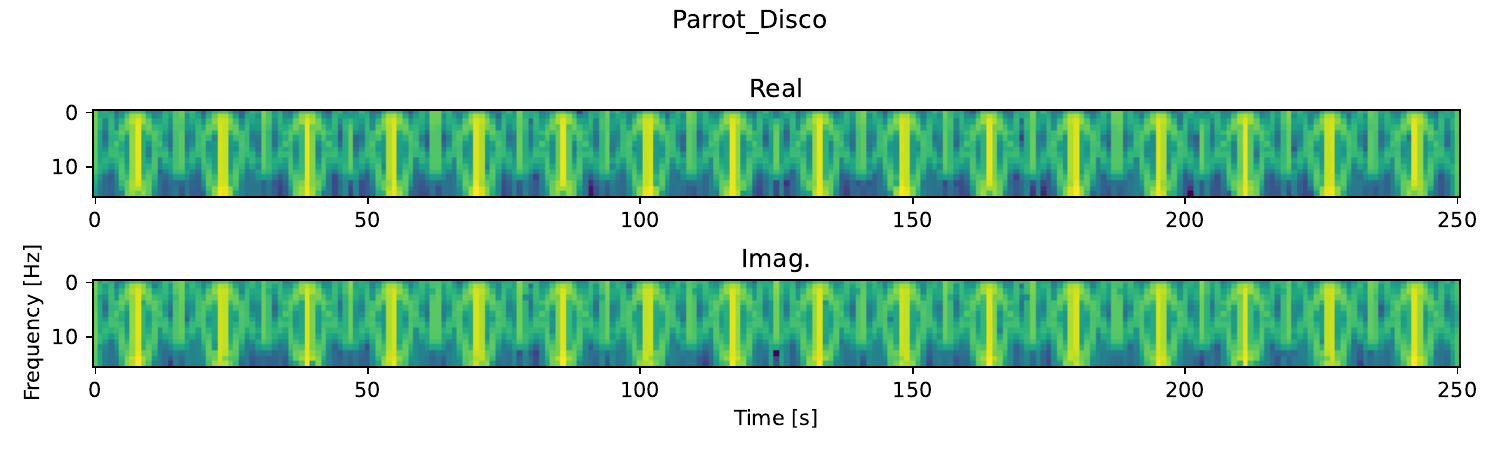}	
	\caption{Real and imaginary valued portions of the STFT of drone Parrot Disco, produced using the MM model. The signal contains no noise.} 
	\label{STFT_drone}%
\end{figure}

\begin{figure}
	\centering 
	\includegraphics[width= .5 \textwidth, angle=0]{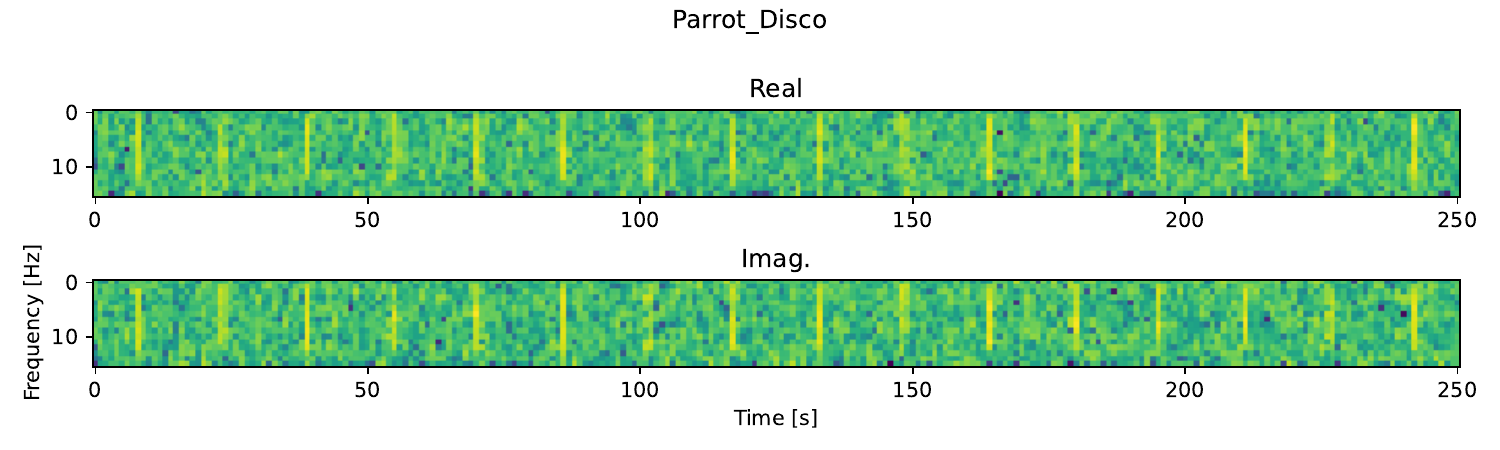}	
	\caption{Real and imaginary valued portions of the STFT for drone Parrot Disco, produced using the MM model, with SNR -5.} 
	\label{STFT_Parrot_SNR-5}%
\end{figure}

\begin{figure}
	\centering 
	\includegraphics[width= .5 \textwidth, angle=0]{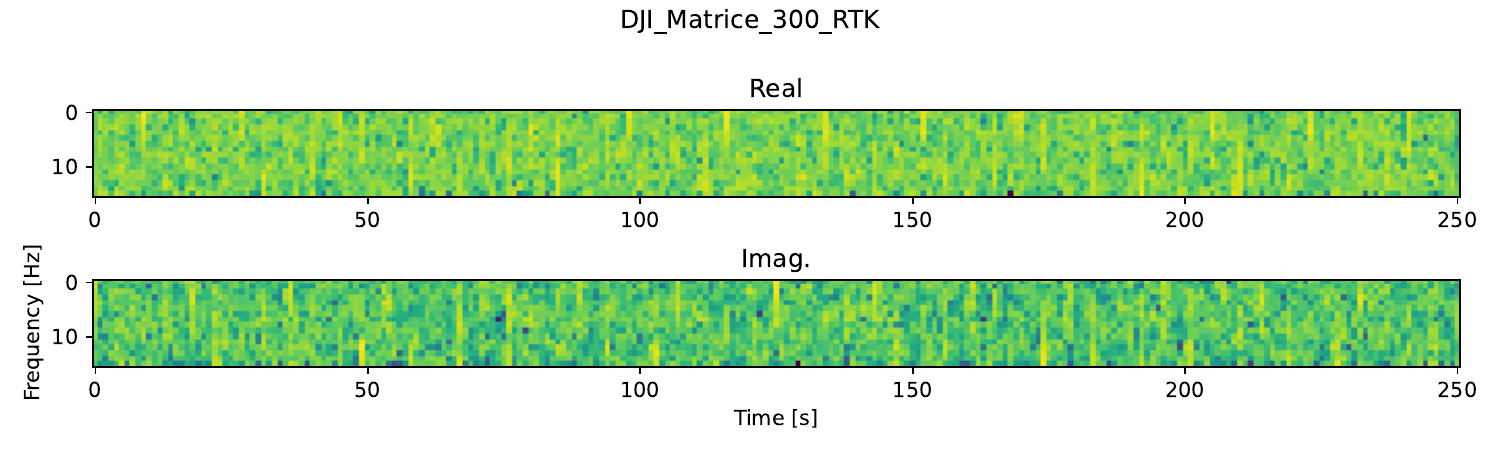}	
	\caption{Real and imaginary valued portion of the STFT for drone DJI Matrice 300 RTK, produced using the MM model, with SNR -5.} 
	\label{STFT_300RTK_SNR-5}%
\end{figure}

\begin{figure}
	\centering 
	\includegraphics[width= .5 \textwidth, angle=0]{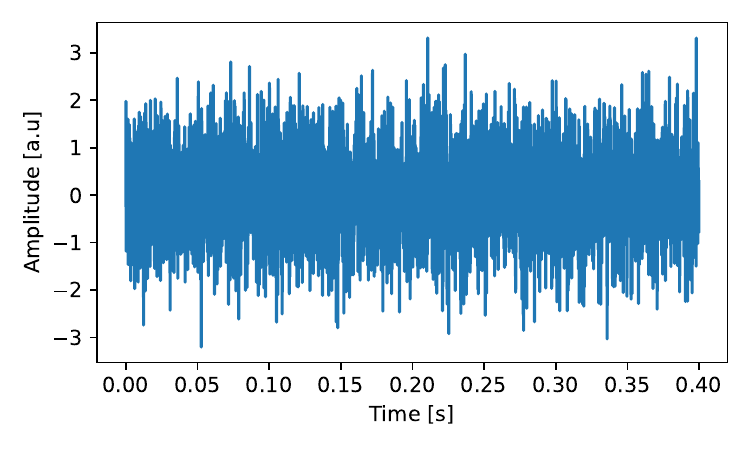}	
	\caption{Time series of gaussian white noise. } 
	\label{time_series_noise}%
\end{figure}

\begin{figure}
	\centering 
	\includegraphics[width= .5 \textwidth, angle=0]{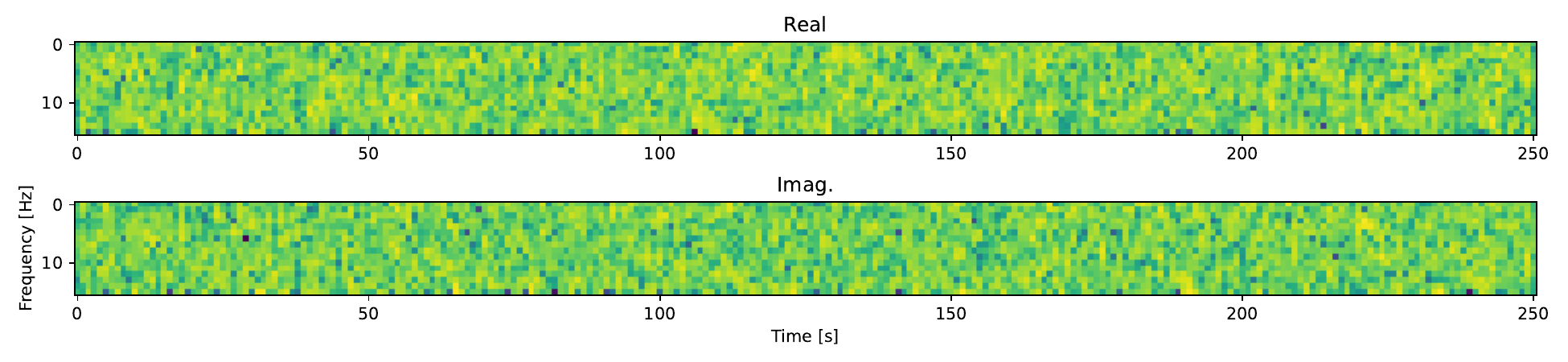}	
	\caption{STFT of gaussian white noise. } 
	\label{STFT_noise}%
\end{figure}

Figure \ref{time_series_drone} shows a time series plot (real and imaginary separately) for a particular drone type, namely the Parrot Disco. The corresponding real and imaginary parts of the STFT of the time domain signal are shown in Figure \ref{STFT_drone}, in the absence of noise. Blade flashes are clearly visible and appear at periodic time intervals, as expected. The STFTs for SNR -5 dB is shown in Fig. \ref{STFT_Parrot_SNR-5}. The blade flashes are less evident, as expected, though there is some residual structure. Figure \ref{STFT_300RTK_SNR-5} shows the STFTs for the same SNR for a different type of drone, namely the DJI Matrice 300 RTK. There are some differences between this figure and the previous one, thus making the utility of ML for detection and classification plausible. Figure \ref{time_series_noise} shows a sample time series of additive Gaussian white noise.  A corresponding STFT produced using noise for both real and imaginary inputs is shown in Figure \ref{STFT_noise}. When the SNR is very low, it is expected that the STFT of any drone will be almost indistinguishable from that of the additive Gaussian noise.  

As shown in prior work, when the time window is short (relative to the blade rotation rate) and the radar PRF is high enough, the spectrogram  reveals the blade flashes. At the other end, when the time window is long, it has been shown analytically (from the Martin-Mulgrew model) as well as experimentally that the resulting STFT has Helicopter Rotor Modulation (HERM) lines\cite{Klaeretal}. In this paper, the short-window spectrograms are utilized in the analysis.

\subsection{Detection and Classification}\label{ssec:DetectionandClassification}

The approach taken in the paper is as follows, which is similar to prior work, such as \cite{spie_conf}). The time domain signal is first converted to a frequency domain signal via an STFT. The separate real and imaginary components of the resulting spectrogram are then used as input to the ML algorithms. 

The regime of interest in practice is the low-SNR regime which is what is studied in detail in this paper.

\section{Simulation Setup}\label{sec:SimulationSetup}

\subsection{Data generation}\label{ssec:DataGen}

We use a short-window Short Time Fourier Transform (STFT) with the two sided spectra and a Hamming window with a window size of 16 samples and window overlap of 8. The sample rate is set to be the same as the radar PRF: 10kHz and the sampling time is 0.2s, sample length is therefore 2000. 

The SNR values of -5 dB to -20 dB, in increments of 5 dB, are used for the detector models. And for the classifier models, SNR values used range between 20 dB to -5 dB, in increments of 5 dB.
We did not explore SNR values lower than -5 dB for the classifier models, as the F1 scores deteriorate rapidly at even lower SNR levels. This can be attributed to the inherent difficulty of drone classification compared to drone detection.

\subsection{Neural Network Architecture}\label{ssec:NNArchitecture}

\subsubsection{CNN}\label{sssec:CNN}

\begin{figure}
	\centering 
	\includegraphics[width= .5 \textwidth, angle=0]{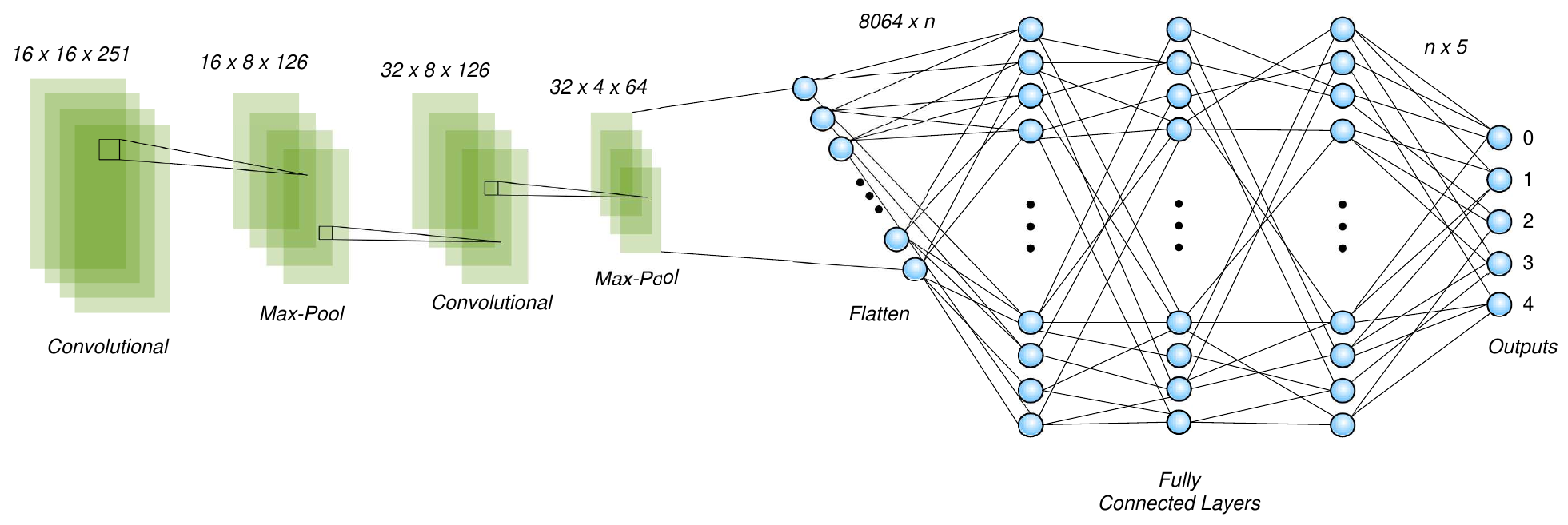}	
	\caption{Architecture of the CNN Classifier for Drone Type Classification. CNN classifier model comprises a convolutional portion followed by a linear portion. Input STFT spectrograms undergo convolutional layers with instance normalization, Leaky ReLU activation, and periodic Max Pooling to extract relevant features. Extracted features are flattened and passed through fully connected layers with Leaky ReLU activation. The final layer produces class probabilities for input radar signals, enabling classification among five drone types.} 
	\label{CNN_architecture}%
\end{figure}

The classical Convolutional Neural Network (CNN) model, illustrated in Fig.\ref{CNN_architecture}, is a two-part architecture consisting of a convolutional portion and a linear portion. The input data, STFT spectrograms, undergo a series of convolutional layers, each followed by instance normalization, Leaky Rectified Linear Unit (Leaky ReLU) activation, and periodic Max Pooling. This process extracts relevant features from the spectrograms. The extracted features are then flattened and passed through three fully connected layers, each with Leaky ReLU activation. The final layer produces the class probabilities for the input radar signals.

To work as a detector and perform binary classification, the final fully connected layer is designed to have two output channels. The model containing this architecture will be referred to as the CNN detector. 

To perform classification among the five drone types, the final fully connected layer is modified to have five output channels. The model containing this architecture will be referred to as the CNN classifier. 

\subsubsection{HQNN}\label{ssssec:HQNN}

\begin{figure}
	\centering 
	\includegraphics[width=0.5\textwidth, angle=0]{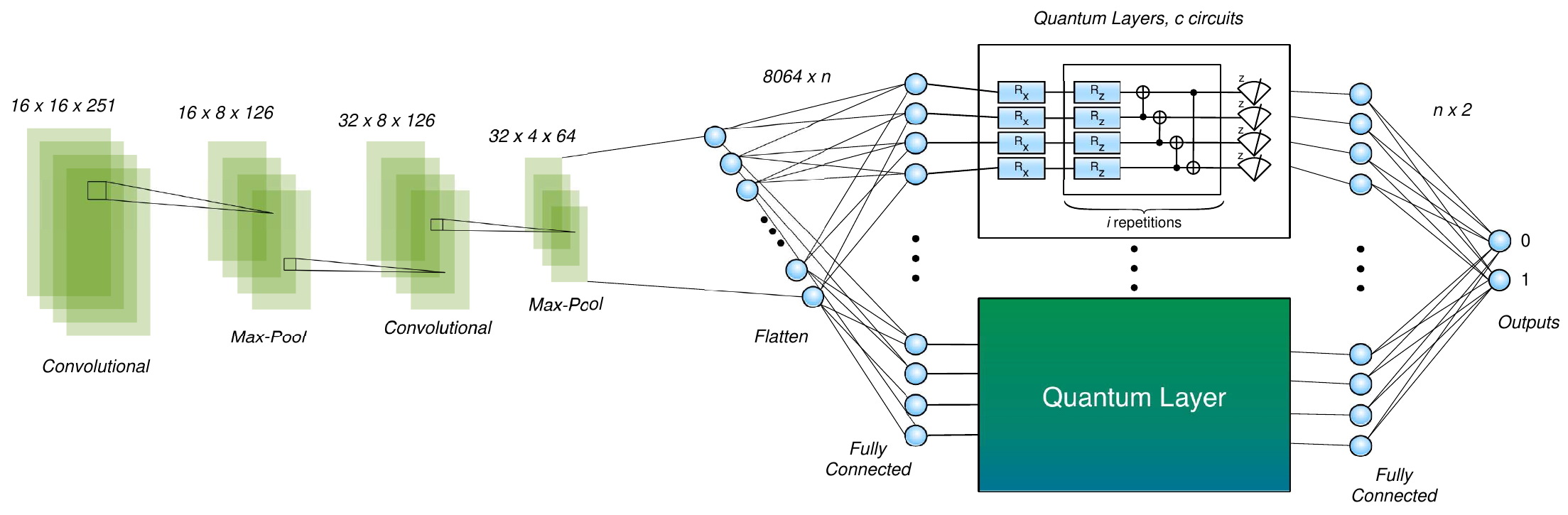}	
	\caption{Architecture of the HQNN detector model. STFT spectrograms undergo convolutional processing for feature extraction and dimensionality reduction. Processed features pass through classical fully connected layers, then split into four tensors for quantum layer processing implemented as VQCs. After quantum processing, tensors are concatenated and refined through a final classical fully connected layer for binary classification. Max pooling and dropout regularization reduce dimensionality and overfitting.} 
	\label{HQNN_architecture}%
\end{figure}

The architecture of the hybrid quantum neural network (HQNN), depicted in Fig. \ref{HQNN_architecture}, employs a hybrid approach that integrates classical convolutional layers with a blend of classical fully connected layers and parallel quantum dense layers. Each quantum layer is implemented as a VQC.

The input to the network, STFT spectrograms, is processed through a sequence of convolutional layers, instance normalization layers, dropout layers, and max pooling layers. This initial convolutional processing extracts and enhances relevant features from the spectrogram data while also reducing the dimensionality.  

The data are then flattened and passed through fully connected layers with Leaky ReLU activation. The processed spectrogram features are then split into four separate tensors, each of which is fed into a separate quantum layer. These quantum layers introduce a non-linear quantum based transformation to the features, potentially capturing more abstract and complex relationships between them. Following the quantum circuit layers, the four processed feature tensors are concatenated and passed through a final fully connected layer with Leaky ReLU activation. These layers further refine and classify the extracted features, leading to the final classification outputs. Max pooling layers are used to reduce the dimensionality of the data. Dropout regularization is applied after the max pooling operations to mitigate over-fitting. 

The quantum part of the HQNN consists of $n$ parallel quantum layers \cite{Hybrid-parallel-model-used}. Each layer is structured as a VQC that encompasses three main components: embedding, variational gates, and measurement. 
The embedding is done using “angle embedding” method. It encodes classical features into the rotation angles of qubits. If the input data for each of the quantum layers is given by the input vector ${x} = (\varphi_{1},\varphi_{2}, \ldots, \varphi_{l}) \in \mathbb{R}^l$, where $l=m/n$ and $m$  represents the features from the previous classical fully connected layer, then the encoding operation is given by $|\psi \rangle = \text{R}_x(x) |\psi_{0}\rangle$ where $|\psi_{0}\rangle = |0\rangle^{\otimes q}$ where $\text{R}_x$ is the rotation gate applied around the X-axis of the Bloch sphere. It rotates each qubit, which is in the ground state, by an angle proportional to the corresponding value in the input vector.
The variational part consists of single-qubit rotations on each qubit with a trainable parameter, followed by a closed chain of CNOT operations. The rotations modify the encoded input data based on the variational parameters while the CNOT gates create entanglement among the qubits. The depth of the variational component $i$ defines the iterations of rotations and CNOT operations within each quantum layer. Variational parameters vary for each repetition $i$ and for individual $c$ quantum circuits. And the total number of weights in the quantum part of the HQNN is given by $l \cdot 3i \cdot c$.
The measurement part consists of the measurement performed on the Pauli basis matrices. The measurement operation results in $v^{(j)} = \langle 0 | R_{x} (\phi_j)^\dagger U(\theta)^\dagger Y_j U(\theta) R_{x}(\phi_j) \left| 0 \right\rangle$
where, $Y_j$ is the Pauli Y matrix for the $j^{th}$ qubit. $R_{x}(\phi_j)$ and $U(\theta)$ are the embedding operation and the operation by the trainable part of the VQC, respectively. And $\theta$ is a vector of trainable parameters.

Following the measurement stage, each quantum layer produces a vector $ v \in \mathbb{R}^l$. The resulting vectors from all the quantum layers are concatenated to form a new vector $\hat{v} \in \mathbb{R}^m$ which is used as the input data for the subsequent classical fully-connected layer.

To work as a detector and perform binary classification, the final fully connected layer is designed to have two output channels. The model containing this architecture will be referred to as the HQNN detector. 

To perform classification among the five drone types, the final fully connected layer is modified to have five output channels. The model containing this architecture will be referred to as the HQNN classifier. 

\subsection{Model Training}\label{ssec:ModelTraining}

HQNN detector and the CNN detector were trained on a dataset of 10,000 drone images (equally among the five drone types) and 10,000 noise images. Model evaluations were performed using a dataset of 5000 drone images (again, equally among the five drone types) and 5000 noise images. 

HQNN classifier and CNN classifier were trained on a dataset of 5000 drone images (equally among the five drone types). Model evaluations were performed using a dataset of 1000 drone images.

We train the CNN detector and the HQNN detector with SNR -5 db data, while we train the CNN classifier and the HQNN classifier with SNR 5 db data. 

Model training runs were executed for at least 50 epochs and achieved a loss threshold of less than $0.005$. All training was done using PyTorch, a Python machine learning library, and Pennylane, a python library for quantum machine learning

\section{Results and Discussion}\label{sec:Results}

 \subsection{Detection Models}\label{ssec:DetectionModels}

\begin{table}[htbp]
\centering
\begin{tabular}{l c c}
\hline
\textbf{SNR} & \textbf{CNN Detector F1 score} & \textbf{HQNN Detector F1 score} \\
\hline

 -5 dB & $0.963 \pm 0.0015275$ & $0.960 \pm 0.003464$ \\
-10 dB & $0.917 \pm 0.0040414$ & $0.946 \pm 0.010263$ \\
-15 dB & $0.824 \pm 0.0183394$ & $0.898 \pm 0.032036$ \\
-20 dB & $0.745 \pm 0.0273922$ & $0.843 \pm 0.043485$ \\

\hline
\end{tabular}
\caption{F1 scores of detection models for various SNR levels. Both CNN detector and HQNN detector are trained with SNR -5 dB data. Standard deviation calculated using three datasets.}
\label{Detection_models_F1_table}
\end{table}

\begin{figure}
	\centering 
	\includegraphics[width=0.5\textwidth, angle=0]{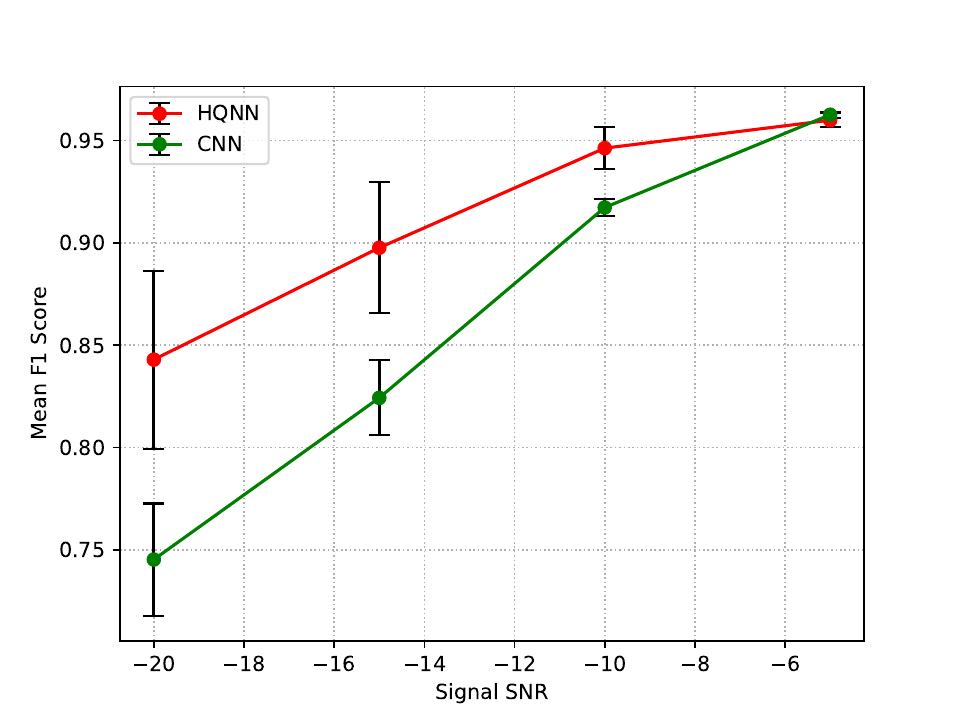}	
	\caption{F1 Score comparison of CNN and HQNN Classifiers trained with SNR 5 dB data. Standard deviation was calculated with three datasets. HQNN detector consistently maintains a higher F1 score compared to the CNN detector at lower SNR levels.} 
\label{F1_CNN_HQNN_detector}%
\end{figure}

We use the F1 score to evaluate trained models, as it accounts for both false positives and false negatives, unlike pure accuracy. The F1 score versus various SNR values of the testing data for the classical and hybrid detection model is shown in Table \ref{Detection_models_F1_table} and plotted in Fig. \ref{F1_CNN_HQNN_detector}. The standard deviation is calculated using three test datasets.

At comparatively higher SNR levels, at -5 dB and -10 dB, both the CNN and HQNN detectors exhibit high F1 scores, with the HQNN detector slightly outperforming. As the SNR decreases, at -15 dB and -20 dB, there is a noticeable decrease in the F1 scores for both detectors. In particular, the HQNN detector consistently maintains a higher F1 score compared to the CNN detector at these lower SNR levels.

\begin{figure}
	\centering 
	\includegraphics[width=0.5\textwidth, angle=0]{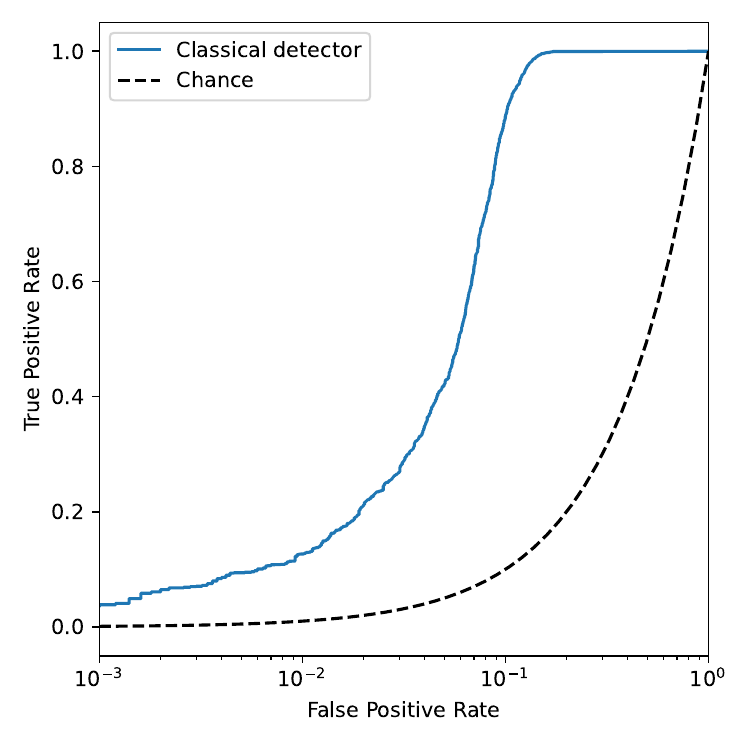}	
	\caption{ROC plot for CNN detector for SNR -20, trained at SNR -5 dB data.} 
\label{CNN_detector_ROC_SNR-5}%
\end{figure}

\begin{figure}
	\centering 
	\includegraphics[width=0.5\textwidth, angle=0]{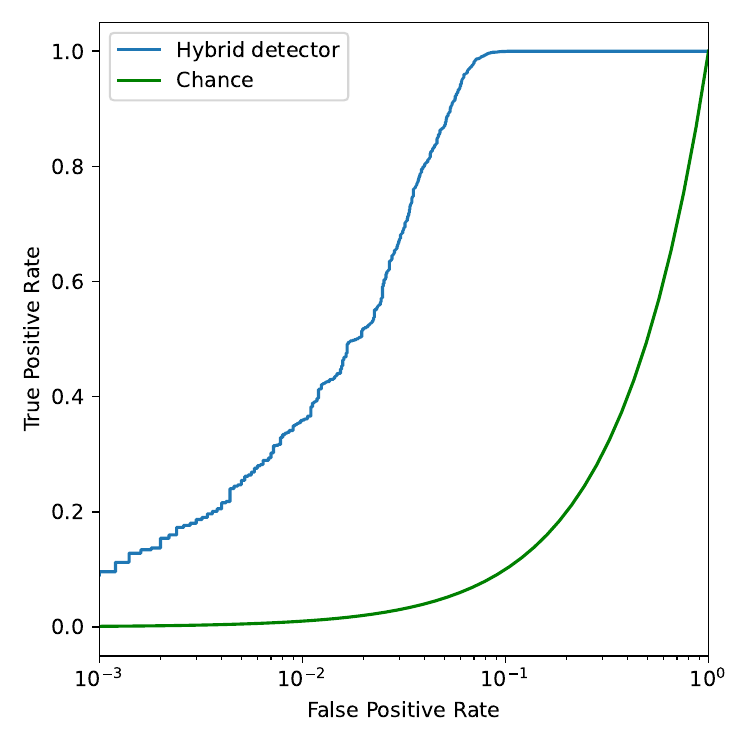}	
	\caption{ROC plot for HQNN detector at SNR -20 dB, trained at SNR -5 dB data.} 
\label{HQNN_detector_ROC_SNR-5}%
\end{figure}

We also use the Receiver Operating Characteristic (ROC) plot to better understand the trade-off between true positives and false positives. This ROC curve is particularly relevant here, as it is standard in radar performance analysis. In addition, we used the log-linear plot for the ROC plot, as is traditional in the radar literature. The ROC plot for the CNN detector for SNR -20 is shown in Fig. \ref{CNN_detector_ROC_SNR-5} and for the HQNN detector in Fig.\ref{HQNN_detector_ROC_SNR-5}.

The noticeable difference between the CNN detector and HQNN detector for the ROC curve can be clearly seen at SNR -20 dB, with the HQNN detector having a better true positive rate value. 

\begin{figure}
	\centering 
	\includegraphics[width=0.5\textwidth, angle=0]{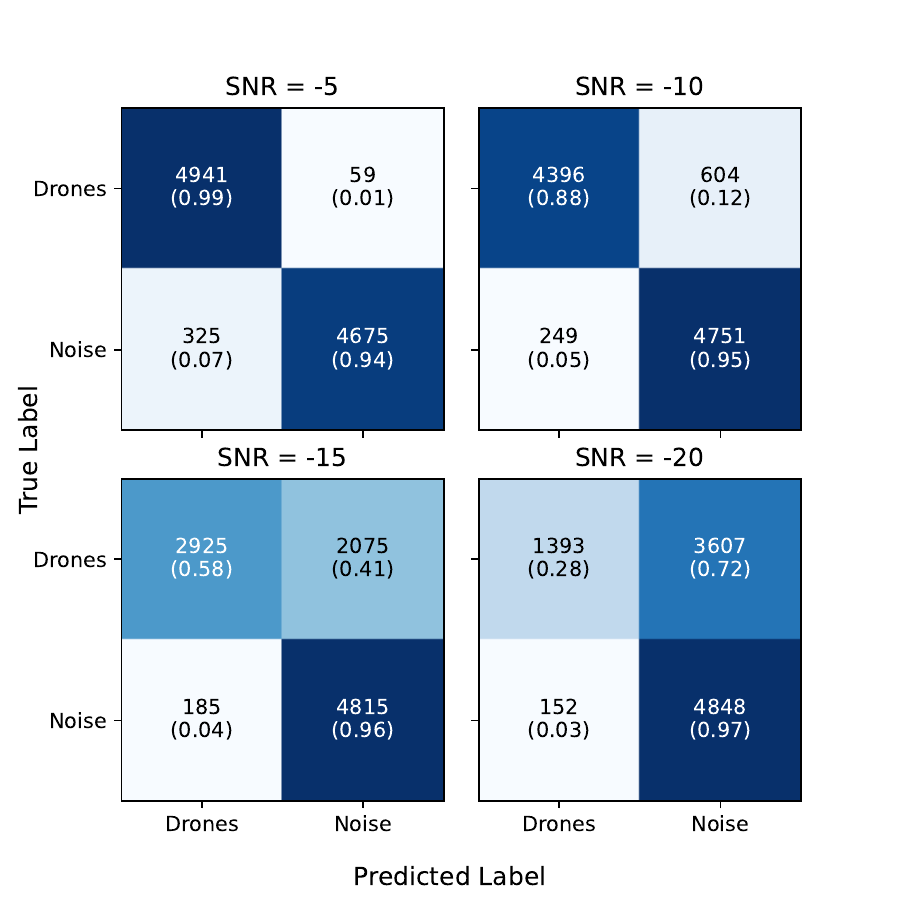}	
	\caption{Confusion matrix of CNN detector trained at SNR -5 dB data.} 
\label{CNN_detector_confusion_matrix_SNR-5}%
\end{figure}

\begin{figure}
	\centering 	\includegraphics[width=0.5\textwidth, angle=0]{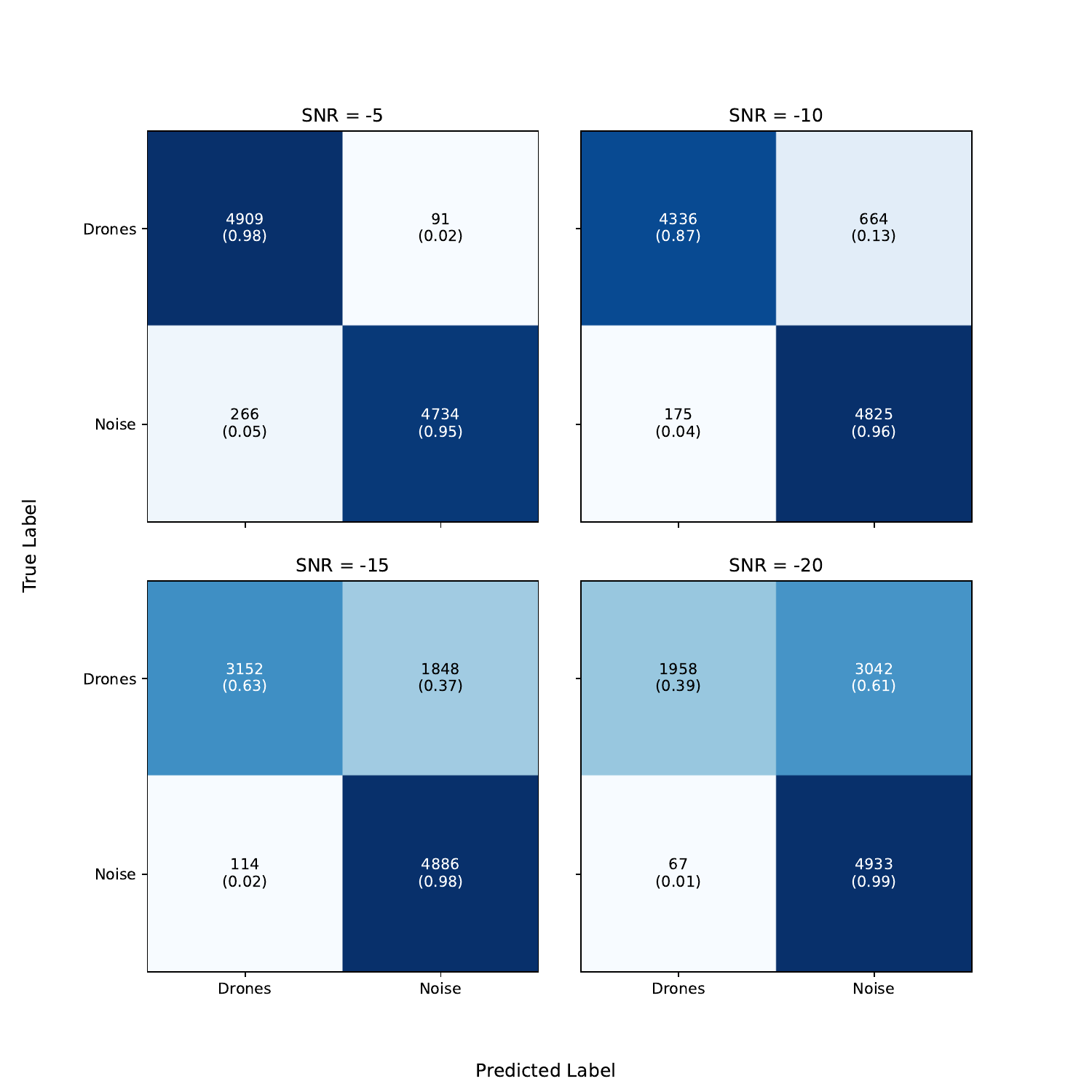}	
	\caption{Confusion matrix of HQNN detector trained at SNR -5 dB data.} 
\label{HQNN_detector_confusion_matrix_SNR-5}%
\end{figure}

We also produce a binary confusion matrix to evaluate the detector models. The confusion matrices for the CNN detector are given in Fig. \ref{CNN_detector_confusion_matrix_SNR-5} and in Fig. \ref{HQNN_detector_confusion_matrix_SNR-5} for the HQNN detector. 
The difference between the detector models becomes evident when examining the confusion matrices across all SNR levels, with particularly notable differences observed at lower SNR levels. At SNR -20 dB, more drones are classified as noise in both models, although the HQNN detector performs better. This is plausible because STFTs at very low SNR values are increasingly indistinguishable from STFT of Gaussian white noise.

\subsection{Classification Models}\label{ssec:ClassificationModels}

\begin{figure}
	\centering 
	\includegraphics[width=0.5\textwidth, angle=0]{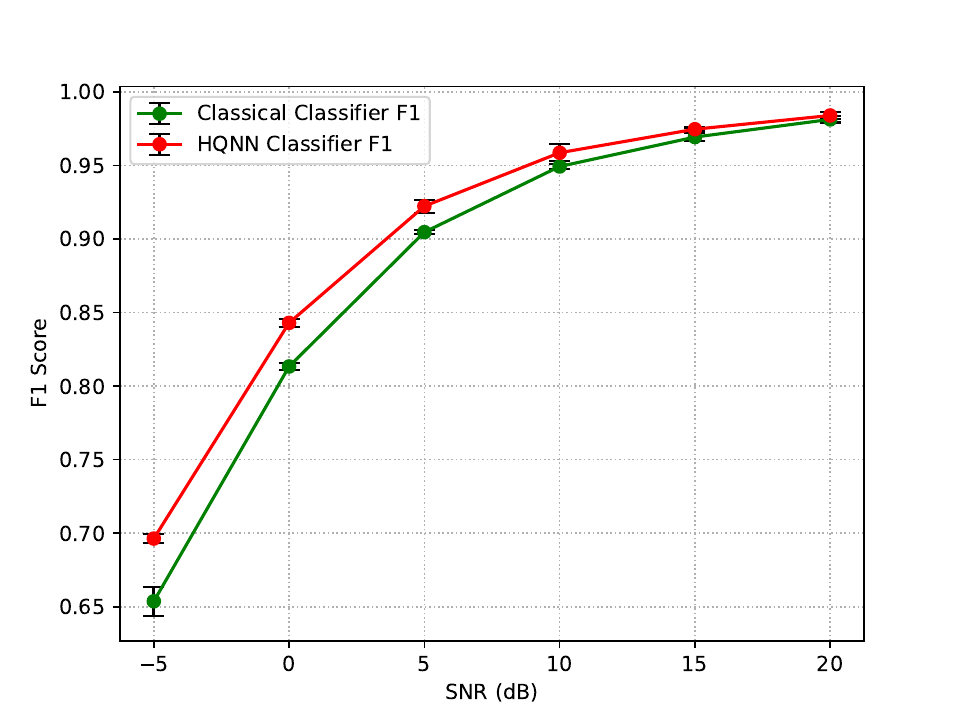}	
	\caption{F1 Score against SNR comparison of CNN and HQNN Classifiers trained with SNR 5 dB data. Standard deviation was calculated with three datasets.} 
\label{F1_CNN_HQNN_classifier}%
\end{figure}

\begin{table}[htbp]
\centering
\begin{tabular}{l c c}
\hline
\textbf{SNR} & \textbf{CNN Classifier F1} & \textbf{HQNN Classifier F1} \\
\hline
20 dB & $0.981 \pm 0.002082$ & $0.984   \pm 0.002646$ \\
15 dB & $0.969 \pm 0.002517$ & $0.975   \pm 0.001155$ \\
10 dB & $0.949 \pm 0.001527$ & $0.959   \pm 0.005686$ \\
5 dB  & $0.905 \pm 0.001527$ & $0.922   \pm 0.004509$ \\
0 dB  & $0.813 \pm 0.002517$ & $0.843   \pm 0.002646$ \\
-5 dB & $0.654 \pm 0.009712$ & $0.696   \pm 0.002887$ \\
\hline
\end{tabular}
\caption{F1 scores of classification models. Both the CNN Classifier and the HQNN Classifier are trained with the SNR 5 dB data set. Standard deviation calculated using three data sets.}
\label{classification_models_F1_table}
\end{table}
For classifer models, SNR values below -5 dB wasn't investigated, as F1 scores degrade rapidly at lower SNR levels, likely because of the greater challenge of drone classification versus detection.
The F1 score for various SNR values of the testing data for the classical and hybrid classification models is shown in Table \ref{classification_models_F1_table} and plotted in Fig. \ref{F1_CNN_HQNN_classifier}. Standard deviation was calculated using three test datasets.
At higher SNR levels, both CNN and HQNN classifiers demonstrate high F1 scores, with the HQNN classifier consistently performing slightly better. As the SNR decreases, there is a noticeable decrease in the F1 scores for both classifiers. However, even at lower SNR levels, the HQNN classifier tends to maintain a higher F1 score compared to the CNN classifier.

\begin{figure}
	\centering 
	\includegraphics[width=0.5\textwidth, angle=0]{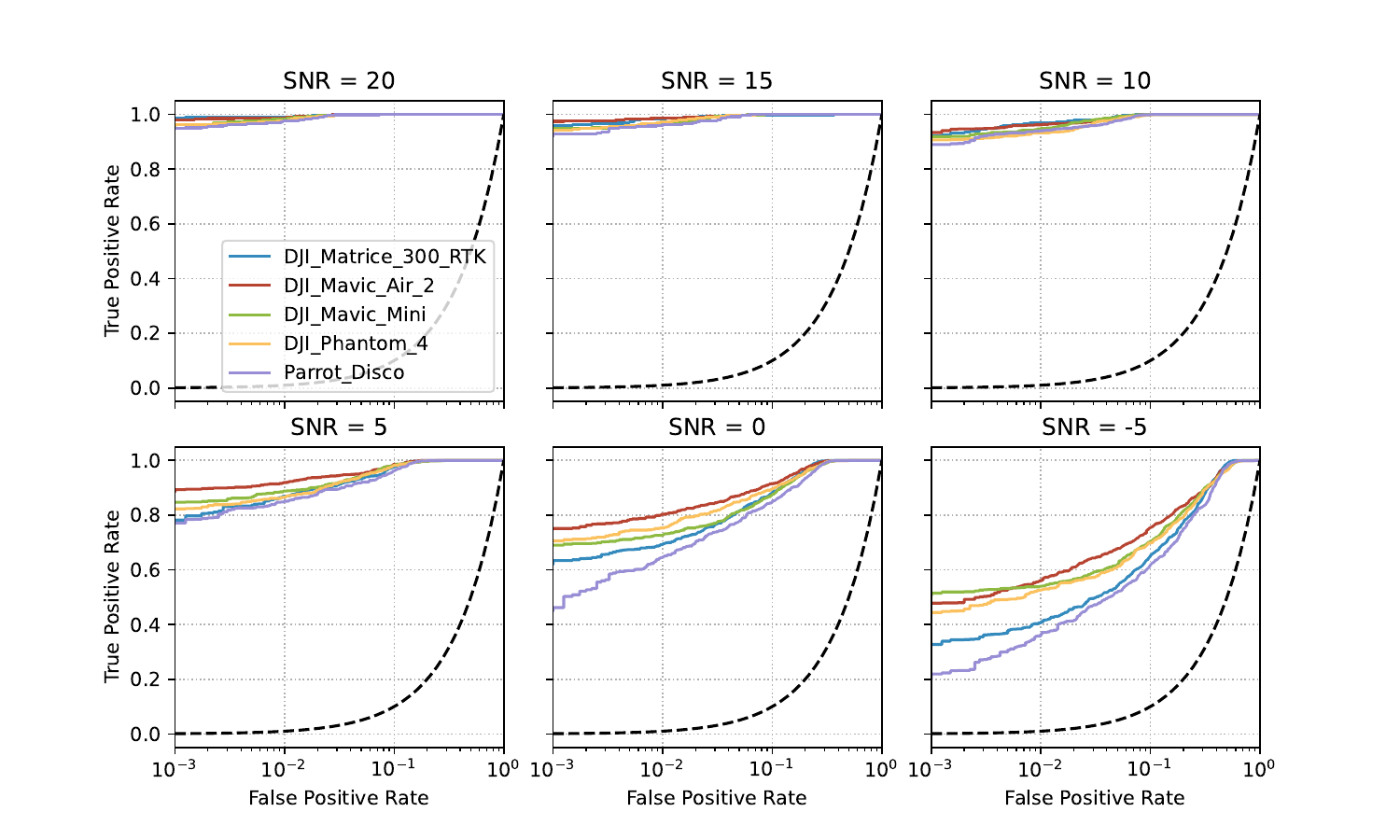}	
	\caption{ROC plots for CNN classifier trained with SNR 5 dB data } 
\label{CNN_classifier_ROC_SNR5}%
\end{figure}

\begin{figure}
	\centering 
	\includegraphics[width=0.5\textwidth, angle=0]{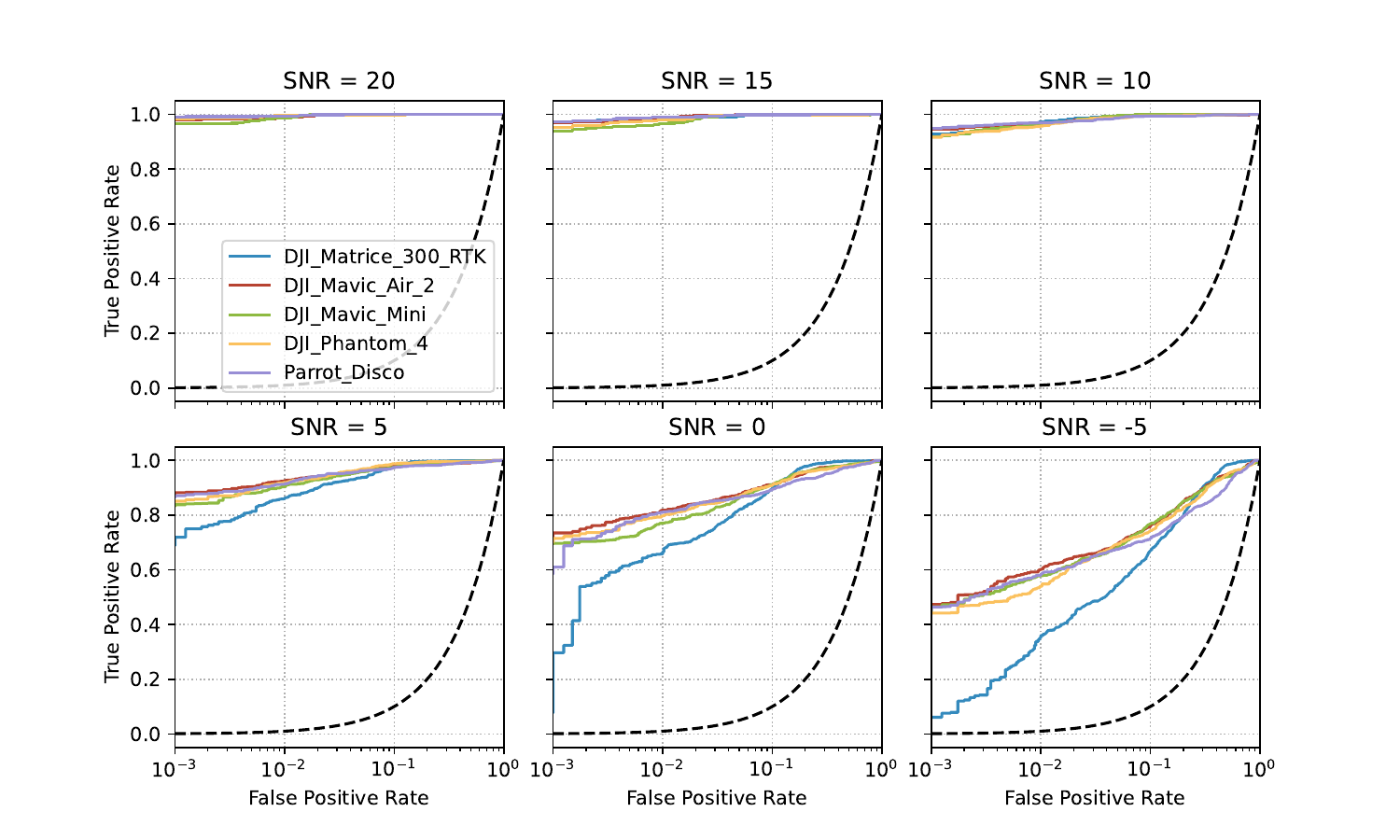}	
	\caption{ROC plots for HQNN classifier trained with SNR 5 dB data} 
\label{HQNN_classifier_ROC_SNR5}%
\end{figure}

We use Multi-class ROC plots to better gain insight into the performance of the CNN classifier and HQNN classifier model. ROC plots for CNN classifier trained with SNR 5 dB data are given in Fig. \ref{CNN_classifier_ROC_SNR5} and for HQNN classifier in Fig. \ref{HQNN_classifier_ROC_SNR5} 

The ROC curves differ with different drone types for both the CNN and HQNN classifier models. The ROC curve for DJI Matrice 300 RTK for the HQNN classifier is considerably worse than for the other drone types and the corresponding performance in the CNN classifier. This could be attributed to DJI Matrice 300 RTK having a very dense micro-Doppler signature (HERM lines) \cite{divy_paper}.
At low SNR levels, for example at SNR -5 dB, the spectrograms of DJI  Matrice 300 RTK (See Fig. \ref{STFT_300RTK_SNR-5}) could exhibit augmentation to a degree where it starts to resemble the STFTs of Gaussian white noise (see Fig. \ref{STFT_noise}), which has no distinct signal. This is in contrast to the STFT of Parrot Disco at SNR -5 dB in Fig. \ref{STFT_Parrot_SNR-5} where the drone signal is more discernible. 
However, the ROC plots for HQNN classifier do perform noticeably better than CNN classifier across all SNR levels for at least some drone types for example, the Parrot Disco. 

\begin{figure}
	\centering 
	\includegraphics[width=0.5\textwidth, angle=0]{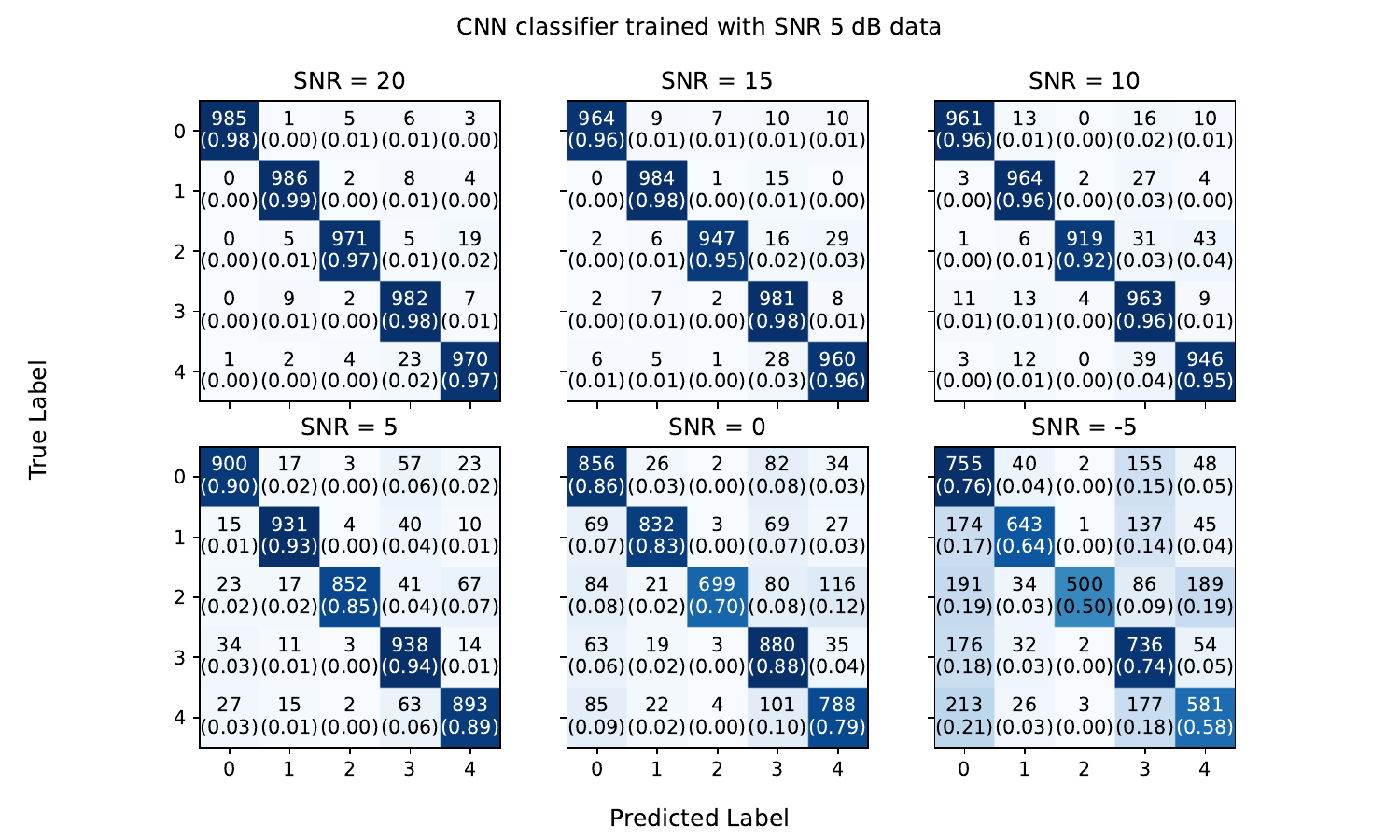}	
	\caption{Confusion matrix of CNN classifier trained at SNR 5 dB data. The labels correspond to the following drone models: 0 for DJI Matrice 300 RTK, 1 for DJI Mavic Air 2, 2 for DJI Mavic Mini, 3 for DJI Phantom 4, and 4 for Parrot Disco. } 
\label{CNN_classifier_confusion_matrix_SNR5}%
\end{figure}

\begin{figure}
	\centering 
	\includegraphics[width=0.5\textwidth, angle=0]{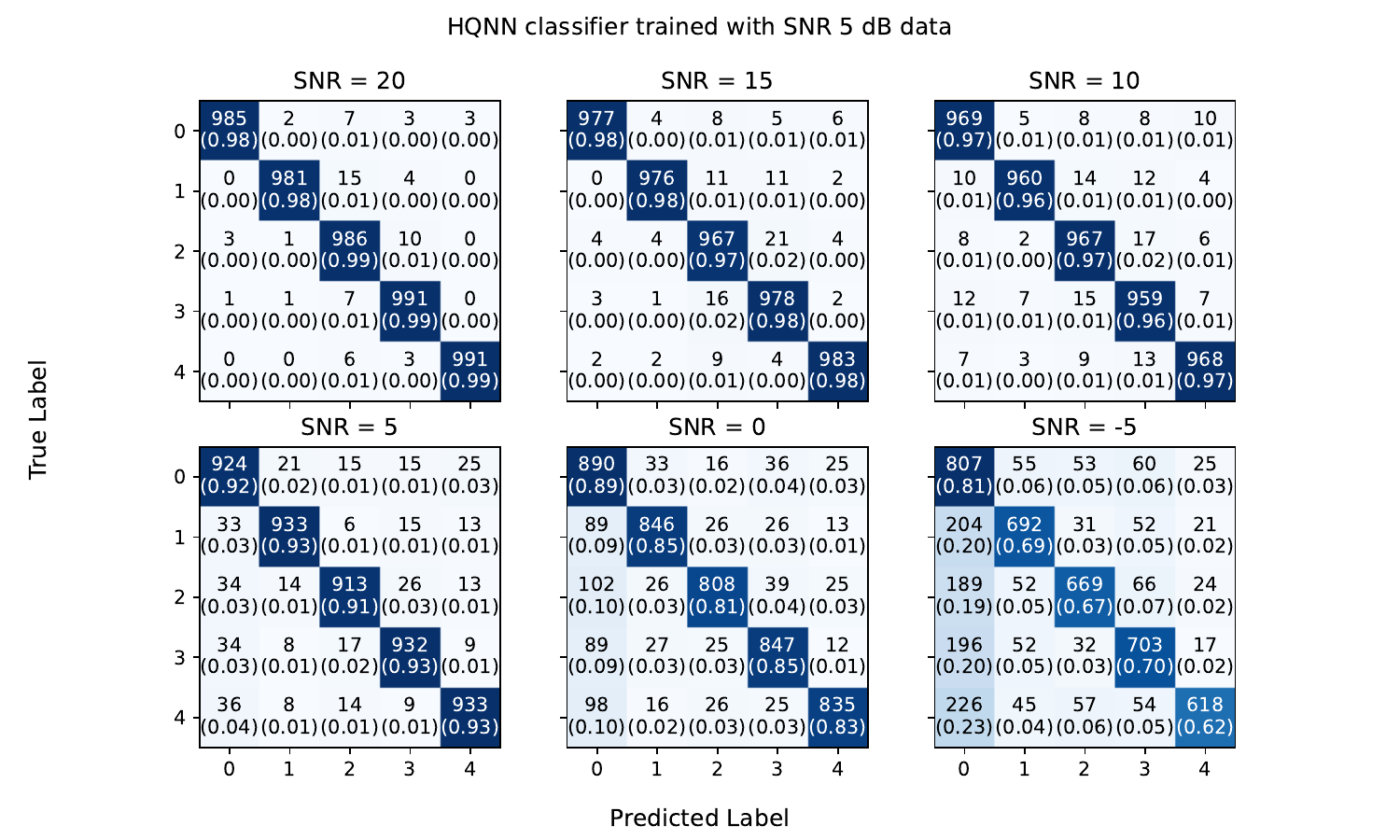}	
	\caption{Confusion matrix of HQNN classifier trained at SNR 5 dB data. The labels correspond to the following drone models: 0 for DJI Matrice 300 RTK, 1 for DJI Mavic Air 2, 2 for DJI Mavic Mini, 3 for DJI Phantom 4, and 4 for Parrot Disco.} 
\label{HQNN_classifier_confusion_matrix_SNR5}%
\end{figure}

We use multi-class confusion matrix to better gain insight into the performance of the classifier models. Confusion matrices for CNN classifier trained at SNR 5 dB data are shown at Fig. \ref{CNN_classifier_confusion_matrix_SNR5} and the confusion matrices for HQNN classifier trained at SNR 5 dB data are shown at Fig. \ref{HQNN_classifier_confusion_matrix_SNR5}
The difference between CNN classifier and HQNN classifier can be especially seen for SNR -5 dB. Both models seem to confuse other drone types for DJI Matrice 300 RTK more often. 

\section{Conclusion and Future Work}\label{sec:Conclusion}

This article investigates the use of HQNNs for the problem of drone detection and classification. We also compare the performance of HQNN with a classical CNN of comparable architecture. 
We observed remarkable detection accuracy, even at extremely low SNR levels of -5, -10, -15, -20, with both classical and hybrid models.
At SNR -20 dB, CNN achieves an F1 score of 0.745 ± 0.0273922 for drone detection, while HQNN outperforms it with an F1 score of 0.843  ± 0.0434856.
At SNR -5 dB, CNN achieves an F1 score of 0.654 ± 0.009712 for drone classification, whereas HQNN achieves a higher F1 score of 0.696 ± 0.002887 for the same task.
Overall, HQNN outperforms a comparable CNN, particularly noticeable at lower SNR levels, highlighting its importance in investigating environments characterized by extremely low SNR levels. 

In future work, one could take advantage of the transfer learning process; the models can be re-trained with real data that may have nuances or variations that are not perfectly captured with the synthetic data. A pre-trained model would allow us to achieve good performance even with limited real data, which is especially advantageous since real data is scarce. Future work could also investigate other possible radar parameters, and also study different neural network architectures, including using a quanvolutional layer, to develop more robust solutions.

This work is preliminary. For example, there is potential for achieving significant detection accuracy even at SNR values lower than -20 dB, when neural network models are trained at appropriate SNR data. A more complicated CNN architecture might outperform HQNN. Nevertheless, it is quite interesting to note that HQNN achieves an unmistakable gain in this problem, and thus motivates the need for further exploration of quantum-inspired algorithms to this important practical problem. 

\section*{Code Availability}
A version of the code used for this work is available in: https://github.com/AishSweety/hybrid-quantum-classical-Neural-Network-for-radar-data

\section*{Conflicts of Interest}
The authors declare no conflict of interest.

\section*{Acknowledgements}

We thank Tim Mathams for the insightful and helpful discussions with signal processing and coding.

\section*{Abbreviations}

\begin{itemize}
\item[]  \textbf{QML}  \sep  Quantum Machine Learning
\item[] \textbf{HQNN}   \sep Hybrid Quantum Neural Network
\item[]  \textbf{STFT}  \sep Short-Time Fourier Transform
\item[] \textbf{MM}  \sep  Martin-Mulgrew 
\item[] \textbf{CNN}   \sep  Convolutional Neural Network
\item[] \textbf{HERM}   \sep HElicopter Rotation Modulation
\item[] \textbf{STFT}   \sep Short-Time Fourier Transform 
\item[] \textbf{SNR}   \sep  Signal-to-Noise Ratio
\item[] \textbf{PRF}   \sep  pulse repetition frequency
\item[] \textbf{ROC}   \sep  Receiver Operating Characteristic
\item[] \textbf{ReLU}  \sep Rectified Linear Unit
\item[] \textbf{VQC}  \sep Variational Quantum Circuit

\end{itemize}

\bibliographystyle{elsarticle-num}

\bibliography{References}
\vspace{12pt}
\end{document}